\documentclass[aps,prl,reprint]{revtex4-2}

\usepackage{amsfonts} 
\usepackage{amsmath}
\usepackage{amsthm}
\usepackage{mathrsfs,xcolor}

\DeclareMathOperator{\Tr}{Tr} 

\begin{document}

\title{On the Spin Dependence of Detection Times\\ and the Nonmeasurability of Arrival Times}

\author{Sheldon Goldstein}
\affiliation{Departments of Mathematics and Physics, Hill Center, Rutgers University,
     110 Frelinghuysen Road, Piscataway, NJ 08854-8019, USA}
\author{Roderich Tumulka}
\affiliation{Fachbereich Mathematik, Eberhard-Karls-Universit\"at T\"ubingen, Auf der Morgenstelle 10, 72076 T\"ubingen, Germany}
\author{Nino Zangh\`{\i}}
\affiliation{Dipartimento di Fisica, Universit\`a di Genova, \\
Via Dodecaneso 33, 16146 Genova, Italy\\
\&\ Istituto Nazionale di Fisica Nucleare (Sezione di Genova) 
}
\date{September 12, 2023}

\begin{abstract}
According to a well-known principle of quantum physics, the statistics of  the outcomes of any quantum experiment are governed by a Positive Operator-Valued Measure (POVM).  In particular,  for experiments designed to measure a  specific  physical quantity,  like the time of a particle's first arrival at a surface,  this principle establishes  that if the probability distribution of that quantity does not arise from a POVM, no such experiment exists. Such is the case with the arrival time distributions proposed by Das and D\"urr \cite{DD19}, due to the nature of their spin dependence.
\end{abstract}

\maketitle
 
 A few years ago, Das and D\"urr, in their work \cite{DD19},   computed  probability distributions for the  times of first arrival of a spin-1/2 particle, moving in an axially symmetric region according to an axially symmetric dynamics, at an axially  symmetric surface  in that region. The calculation was conducted within the framework of Bohmian mechanics, which provides a well-defined concept of arrival time, in contrast to standard quantum mechanics.

The probability distributions they found depend upon the spin of the particle and  for some choices of the spin the distribution exhibits  intriguing characteristics.  They assumed that the arrival times could be measured using appropriate detectors. In fact, it was partly on the basis of the nature of the spin dependence that they argued for this assumption. 

However, on the basis of the spin dependence,  we prove here that their assumption regarding the measurability of arrival times is false: there can exist no such detectors. This conclusion is grounded in a well-known principle: The statistics of  the outcomes of any quantum experiment are governed by a Positive Operator-Valued Measure (POVM). This principle is considered a fundamental tenet in standard quantum mechanics, as exemplified in references such as \cite{Davies}. Moreover,  it is  a theorem within the framework of Bohmian mechanics \cite{DGZ04}.

 In the specific context of an experiment designed to measure a particular physical quantity,  like the particle's  time of first arrival at a surface,  this principle establishes  that if the probability distribution of that quantity does not arise from a POVM, no such experiment exists. As we show, such is the case with the arrival time distributions found in \cite{DD19}, due to the nature of their spin dependence.

\paragraph{Necessary condition on spin dependence.---}Consider an experimental setup involving a spin-1/2 particle.  The outcome statistics  is given by a POVM $\mathcal{O}= \mathcal{O}(dt)$, i.e., when 
the quantum state of the particle  is $\Psi$,  the distribution of results can be expressed as $\langle \Psi | \mathcal{O} | \Psi \rangle$.

Now, let us introduce ${\bf n}$, an arbitrary unit vector in three-dimensional space. A spinor $|{\bf n}\rangle$ in $\mathbb{C}^2$ 
is defined uniquely up to a phase by the condition
$ {\bf n}  = \langle {\bf n} |  \boldsymbol{\sigma} |{\bf n}\rangle
$,
where $\boldsymbol{\sigma}=(\sigma_x, \sigma_y, \sigma_z) $ constitutes the vector of Pauli matrices. 

Suppose that $\Psi$ takes the form of a product state, 
\begin{equation}\label{product}
\Psi = \psi \otimes |{\bf n}\rangle,
\end{equation}
where $\psi$ accounts for the translational degrees of freedom. When considering $\psi$ as fixed and the spin state $|{\bf n}\rangle$ as variable, the spin-dependent aspect of the result's statistics can be encapsulated by the {\em spin POVM}  $O=O(dt) =\langle\psi|\mathcal{O}(dt)|\psi\rangle $. 
Consequently, the statistics of the result for the state defined in equation \eqref{product} can be expressed as  
\begin{equation}\label{POVM}
 P_{\bf n}  = P_{\bf n}(dt)   = \langle {\bf n} |  O (dt)  |{\bf n}\rangle.
\end{equation}
This severely constrains  the possible spin dependence of the outcome statistics $P_{\bf n}$.
Since $|{\bf n}\rangle$ and $|{\bf -n}\rangle$ are orthogonal, we have that
\begin{equation}\label{trace}
P_{\bf n} + P_{\bf -n}=\Tr(O),
\end{equation}
so that it  does not depend on the chosen direction $\bf n$ of the spin vector. We thus arrive at the following  necessary condition:
  \begin{equation}\label{eq:NC} \begin{minipage}{.42\textwidth}{\it Let $\mathscr{P}_{\bf n}=\mathscr{P}_{\bf n} (dt)$ be any spin dependent distribution, for example that  of a physical quantity such as an arrival time  for a system in the state \eqref{product}. For this to arise from a spin POVM it must be the case that  
$\mathscr{P}_{\bf n} + \mathscr{P}_{\bf -n} $ does not depend on the chosen direction $\bf n$ of the spin vector. } \end{minipage}\end{equation}

\paragraph{Axial symmetry.---}Assuming axial symmetry, let us consider a unit vector ${\bf a}$ aligned with the symmetry axis and another unit vector ${\bf b}$ perpendicular to ${\bf a}$. We write  $P_{\uparrow}$ for $P_{\bf a}$,  $P_{\downarrow}$ for $P_{-\bf a}$,  and $P_{\rightarrow}$ for $P_{\bf b}$, with similar notation for $\mathscr{P}$. Axial symmetry implies that $\mathscr{P}_{\bf b}=\mathscr{P}_{-\bf b}$. Thus, choosing $\bf a$ and $\bf b$ for $\bf n$ in \eqref{eq:NC}, we have  that
\begin{equation}\label{DTPE}
\begin{minipage}{.42\textwidth}
{\it Under axial symmetry, the spin-dependent distribution $\mathscr{P}_{\bf n}$ of a physical quantity in the state \eqref{product} can arise from a spin POVM only if 
$$\mathscr{P}_{\uparrow} + \mathscr{P}_{\downarrow} = 2\mathscr{P}_\rightarrow. $$ } \end{minipage}\end{equation} 

Suppose now that, in addition to axial symmetry, we have also that 
 $\mathscr{P}_{\uparrow} = \mathscr{P}_{\downarrow}$. In this case  \eqref{DTPE} becomes the following:
\begin{equation}\label{rm1}
\begin{minipage}{.42\textwidth}
{\it Under axial symmetry, if $\mathscr{P}_{\uparrow} = \mathscr{P}_{\downarrow}$, then  $\mathscr{P}_{\bf n}$  can arise from a spin POVM only if $\mathscr{P}_{\uparrow}= \mathscr{P}_{\rightarrow}$.}
\end{minipage}
\end{equation}

\paragraph{Arrival times in Bohmian mechanics.---} 
 In the context of Bohmian mechanics, arrival times are well-defined, and the crucial question revolves around their measurability. In   \cite{DD19},  Das and D\"urr calculated the arrival time distributions $\mathscr{P}_{\bf n}$ and found intriguing spin-dependent characteristics. In addition to having axial symmetry, the arrival time distributions  were such that $ \mathscr{P}_{\uparrow}= \mathscr{P}_{\downarrow}$. Moreover,
they  found that $\mathscr{P}_\uparrow \neq \mathscr{P}_\rightarrow$ (in fact, dramatically so). Thus, by \eqref{rm1}, the arrival times distributions don't arise from a POVM, and hence the arrival times are not measurable.

\paragraph{Approximate measurement.---} If a physical quantity can't be measured with arbitrary precision, it still might be possible to perform an approximate measurement of the quantity. It is natural to then ask, given  $\mathscr{P}_{\bf n}$, how closely this can be approximated by a spin POVM. We have no definitive answer to this question, but we can provide a simple lower bound on the size of the error.  

Let\begin{equation}\label{Delta}
    \Delta= \sup_{{\bf n},{\bf m}}\|(\mathscr{P}_{\bf n} + \mathscr{P}_{\bf -n})-(\mathscr{P}_{\bf m} + \mathscr{P}_{\bf -m})\|,
\end{equation}
where $\|\cdot\|$ denotes, say, the total variation norm---$L^1$ for absolutely continuous measures. Then it is easy to see \cite{fn1} that any approximating spin POVM must yield an error of at least $\Delta/4$ (or $\Delta/4-\epsilon$ if the continuity of $\mathscr{P}_{\bf n}$ as a function of $\bf n$ is not assumed), in the sense that for any (approximating) spin POVM $O$, there must be a direction $\bf n$ such that $\|\mathscr{P}_{\bf n}- \langle {\bf n} |  O  |{\bf n}\rangle\|\ge \Delta/4-\epsilon$. 
Thus for any POVM $O$, with associated probability distributions $P_{\bf n}$, we have that
\begin{equation}\label{approx}
 \|\mathscr{P}-P\|\equiv \sup_{\bf n} \|\mathscr{P}_{\bf n}- P_{\bf n}\|\ge \Delta/4. 
\end{equation}

As a consequence, for the model of \cite{DD19}, in which $\mathscr{P}_{\uparrow} = \mathscr{P}_{\downarrow}$, any potential arrival time {\it measurement} must involve an error of at 
 least $\|\mathscr{P}_{\rightarrow}-\mathscr{P}_{\uparrow}\|/2$ in the above sense. Since this quantity is not negligible, the family of the $\mathscr{P}_{\bf n}$ can't be close to any family of $P_{\bf n}$, contrary to the expectation of Das and D\"urr that the presence of detectors would cause only a mild disturbance \cite{DD19}.

\paragraph{Inversion symmetry.---}In addition to obeying $ \mathscr{P}_{\uparrow}= \mathscr{P}_{\downarrow}$, the arrival time distributions $\mathscr{P}_{\bf n}$ found in \cite{DD19} have full inversion symmetry, obeying 
$\mathscr{P}_{\bf n}=\mathscr{P}_{\bf -n}$ for all $\bf n$. It  follows from \eqref{eq:NC} that under this {\it axial-inversion} symmetry, the arrival time distributions, if measurable, could involve no spin dependence whatsoever. 

\paragraph{Chiral symmetry.---}In fact, a more detailed examination of the form of an axially symmetric spin POVM \cite{fn2}  shows that, assuming only axial symmetry, $P_{\bf n}$  can have no spin dependence at all if it obeys $P_{\uparrow}=P_{\downarrow}$, an expression of chiral symmetry,  i.e., symmetry between the clockwise and the counter-clockwise sense of rotation about the ${\bf a}$-axis. Thus any detector that respects the axial and chiral symmetry of the model in  \cite{DD19}, regardless of how accurate or inaccurate it may be, will find no spin dependence for {\it detection} time statistics. In particular such detectors will reveal no difference in what they find for directions in which the {\it arrival} time distributions have intriguing characteristics and those in which they don't.


\begin{thebibliography}{100}
\bibitem{fn1} Indeed, by \eqref{Delta} there are $\bf n$ and $\bf m$ such that the norm in \eqref{Delta} is within $4\epsilon$ of the supremum $\Delta$.  By \eqref{trace}, replacing $\mathscr{P}_{\bf n}$ in \eqref{Delta} by $\mathscr{P}_{\bf n}-\langle {\bf n}|O|{\bf n}\rangle$, and likewise for $-{\bf n},{\bf m},-{\bf m}$, leaves the norm in \eqref{Delta} unchanged. By the triangle inequality, this norm is $\leq$ the sum of $\|\mathscr{P}_{\bf n}-\langle {\bf n}|O|{\bf n}\rangle\|$ and the corresponding terms for $-{\bf n},{\bf m},-{\bf m}$; thus, not all of the 4 summands can be $< \Delta/4-\epsilon$.
\end{thebibliography}
\end{document}